**Substitutional mechanism for growth of hexagonal boron nitride on epitaxial graphene**

Patrick C. Mende, Jun Li, and Randall M. Feenstra[*]
Dept. Physics, Carnegie Mellon University, Pittsburgh, PA 15213

**Abstract**
Monolayer-thick hexagonal boron nitride (h-BN) is grown on graphene on SiC(0001), by exposure of the graphene to borazine, $(BH)_3(NH)_3$, at 1100 °C. The h-BN films form ~2-μm size grains with a preferred orientation of 30° relative to the surface graphene. Low-energy electron microscopy is employed to provide definitive signatures of the number and composition of two-dimensional (2D) planes across the surface. These grains are found to form by substitution for the surface graphene, with the C atoms produced by this substitution then being incorporated *below* the h-BN (at the interface between the existing graphene and the SiC) to form a new graphene plane.

The vertical stacking of 2D materials creates many possibilities for electronic devices.[1] Considering just hexagonal boron nitride (h-BN) and graphene, h-BN has been shown to be both an ideal substrate on which to fabricate graphene-based devices and a very useful encapsulation layer for such devices.[2,3] Furthermore, h-BN is known to be an excellent dielectric layer to serve as a gate insulator or a barrier layer in a tunneling device. A number of groups have demonstrated graphene-insulator-graphene (GIG) devices utilizing h-BN barrier layers, but in all cases such devices have been prepared either by exfoliation from bulk crystals or by chemical-vapor deposition (CVD) on metal foils.[4,5,6,7] Complete hetero-epitaxial growth of the h-BN on graphene, without the need for transfer, is preferable from the point of view of both large-area growth (for scalable device fabrication) and to avoid the processing steps (and concomitant possible contamination) associated with exfoliation and transfer.

In this work, we study the heteroepitaxial growth of h-BN on graphene, with the graphene itself being so-called *epitaxial graphene* (EG) formed on SiC. Use of EG, as opposed to e.g. graphene formed on a metal substrate such as Cu or Ni,[8,9,10,11] offers the advantage of an insulating substrate (i.e. utilizing semi-insulating SiC) for electronic devices. Whereas the growth of h-BN on catalytic metal substrates has been studied by many groups, as recently reviewed by Yin et al.,[12,13] the mechanism for h-BN growth directly on graphene (or EG) without any underlying metal is a less well understood topic.[14,15,16,17,18]

Our growth of h-BN is accomplished by exposing the EG to borazine, $(BH)_3(NH)_3$, at a pressure of $1\times10^{-4}$ Torr and temperatures in the range 950 – 1300 °C (see Supplementary Material for further details);[19,20] we focus here on results obtained at 1100 °C. Characterization of the h-BN is performed *in-situ* using wide-area low-energy electron diffraction (LEED), both before and after growth, and *ex-situ* using atomic force microscopy (AFM), low-energy electron microscopy (LEEM) including selected-area μLEED (1.25 μm spatial resolution), and low-energy electron reflectivity (LEER).

---

[*] feenstra@cmu.edu

Figure 1 shows how the LEED pattern of the sample evolves with the growth of h-BN. Figure 1(a), acquired before the h-BN growth, is typical of epitaxial graphene on SiC(0001), with SiC, graphene, and the buffer-layer satellite spots readily apparent[21,22,23,24] (see Fig. 3(d) of Ref. [25] for a LEED pattern of the buffer layer only). The relative intensities of the graphene and SiC spots in Fig. 1(a) is consistent with graphene which is between 1 and 2 MLs thick.[24,25] The post-growth pattern in Fig. 1(b) shows an additional faint ring of intensity, with wavevector very nearly equal to that of the primary graphene spots. Careful measurement of the wavevectors of the graphene and h-BN spot positions (on a separate LEED pattern on which they are better resolved; see Fig. S1 of the Supplementary Material) reveals that the average wavevector magnitude of the latter is 1.89 ± 0.13% smaller than the former, consistent with the 1.8% larger lattice parameter of h-BN compared to graphene. Hence, we find that unstrained h-BN is forming on the surface, with preferential orientation of the h-BN grains at 30° to the graphene orientation.

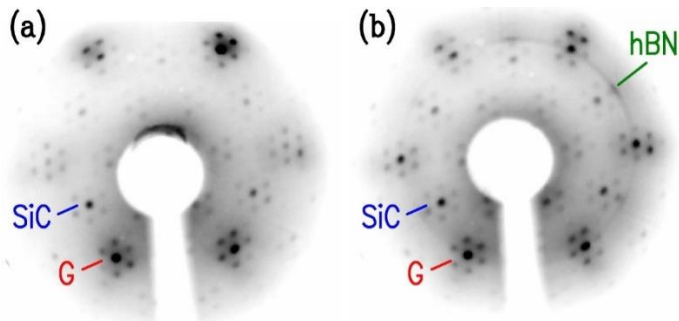

FIG 1. LEED patterns acquired at 100 eV, from (a) epitaxial graphene (EG) on SiC, and (b) after exposure to $1\times10^{-4}$ Torr borazine at temperature of 1100 °C. The SiC and graphene (G) spots are indicated (each of them surrounded by six satellite spots). A ring of intensity arising from h-BN is also apparent (with broad maximum at the location indicated).

More detailed characterization of the h-BN films is obtained using LEEM, as shown in Fig. 2. Figure 2(a) shows a LEEM image with a 10 V "starting voltage" (this is the voltage bias of the sample relative to the electron emitter, which corresponds approximately to the incident electron energy relative to the vacuum level, aside from a small correction for the work function difference between emitter and sample which we quantify in Fig. S2).[26] At this voltage, a few isolated dark triangles are readily apparent. As determined in our prior work,[27] this energy range corresponds to a reflectivity minimum specifically for h-BN but not for graphene, and hence we can identify the ~2-μm-size triangles seen in Fig. 2(b) as h-BN islands. Figure 2(b) shows a specific area of Fig. 2(a), with one h-BN island highlighted by white dashed lines. LEER spectra are acquired from the points labeled A – D, as shown in Fig. 2(c); these spectra provide a powerful "fingerprinting" tool,[28] enabling us to discriminate between different combinations of h-BN and graphene.[26,27,29,30]

Specifically, in the low-energy range of 0 – 6 eV, the oscillations in the LEER spectra arise from *interlayer states* of the h-BN and graphene,[31,32,33] and by counting the number of these minima we can determine the number of spaces between h-BN and/or graphene planes. Importantly, the graphene-like *buffer layer* that exists below the EG,[34,35,36,37,38,39,40] often referred to as layer $G_0$, does not have any interlayer state between it and the SiC.[31] Hence, for the first EG layer, which we refer to as $G_1$, there is a single interlayer state (between $G_0$ and $G_1$) and a single minimum in the reflectivity. Curve A of Fig. 2(c) shows this type of spectrum. With additional graphene or h-BN layers above $G_1$, then additional minima appear in this 0 – 6 eV range of the spectra. For the broad minimum seen in some of the LEER spectra at 8 – 12 eV, this feature has been definitively identified as being derived from states of the h-BN that are localized *on* the atomic planes.[27] Such *intralayer states* are forbidden, by symmetry, to produce reflectivity minima for graphene.[27] Hence, as already mentioned above, this minimum at ~10 eV allows us to identify the presence of h-BN. Thus, we can identify the number of graphene and h-BN layers associated with each of the spectra in Fig. 2(c), as labelled in the figure (additional LEEM/LEER data are presented in Fig. S3). Auger electron spectroscopy data also reveals the B and N peaks arising from the BN islands on the surface, both for the present samples and for sample of BN deposited on Cu utilizing the same borazine deposition system.[26]

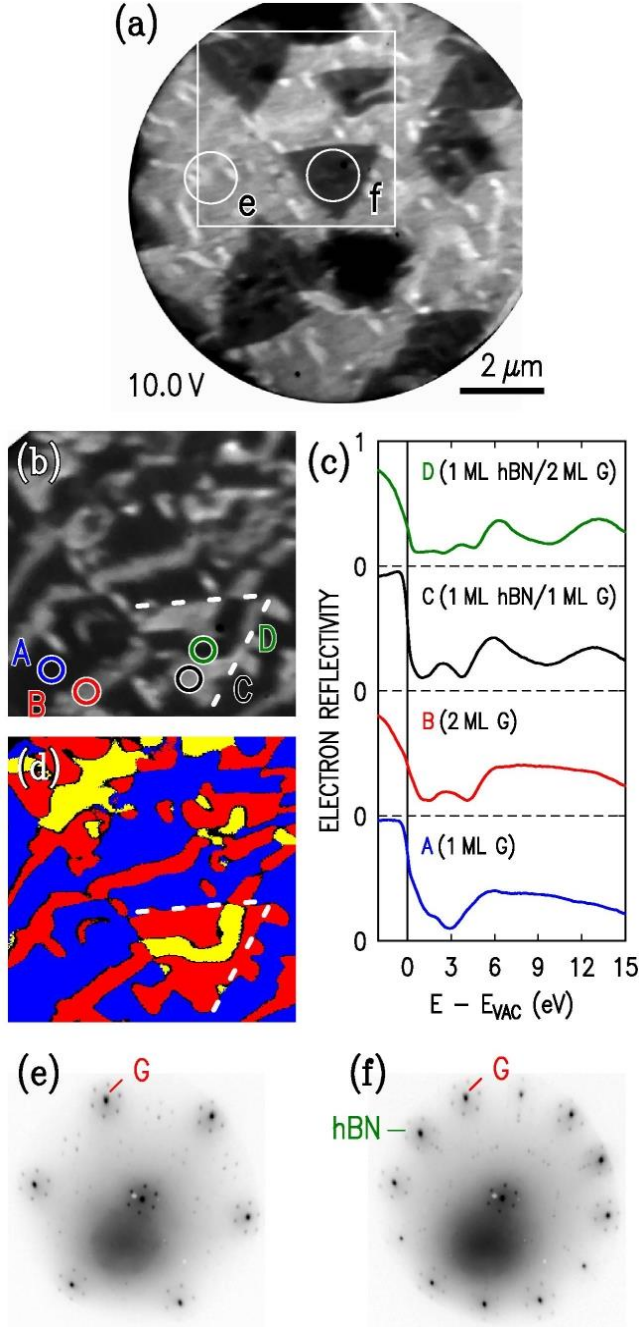

FIG 2. (a) LEEM image of sample grown at 1100°C, acquired at 10 V start voltage and showing many h-BN islands in the field of view. (d) LEEM image from area indicated in (a), but with a start voltage of 3.8 V corresponding to a LEER minimum for both 1 and 3 MLs of 2D material. A dashed triangle indicates the h-BN island in (a) which contains the circle labeled "c". (c) LEER spectra, acquired from the locations marked in (b). (d) Thickness map of the same area as in (b). Blue, red, and yellow regions correspond to 1, 2, and 3 MLs of graphene/h-BN, respectively. (e) and (f) μLEED patterns from the areas indicated in (a). Graphene and h-BN spots are labeled.

To more clearly illustrate the number of 2D layers at all locations on the surface, we perform a thickness mapping using a method described elsewhere,[36] as presented in Fig. 2(d). It can be clearly seen that the highlighted h-BN island contains 1 more 2D plane compared to the area around the island. Most of the island consists of 1 ML of h-BN plus a graphene layer, but there is a strip of material (yellow in Fig. 2(d)) extending through the island for which there are 2 graphene layers below the h-BN. Similarly, around the h-BN island there is typically a single graphene layer, but some areas (such as the red strip connected to the yellow one that extends through the h-BN) there are 2 graphene layers. As shown by AFM data below, these "strips" of material occur frequently on EG, forming near step edges. We find that the h-BN islands tend to form near the step edges (i.e. heterogeneous nucleation), as illustrated in Fig. S4.

Selected-area diffraction (µLEED) results are presented in Figs. 2(e) and 2(f), acquired from the surface areas indicated in Fig. 2(a). Inspecting Fig. 2(e), obtained from a region of the surface that contains only graphene, at the center of the pattern is the specularly reflected (0,0) beam which is surrounded by the six satellite spots of the 6√3 buffer-layer structure.[47] For the diffraction pattern of Fig. 2(f), which comes from a single h-BN island as seen from the LEEM image, we see the same primary SiC and graphene spots as those from the area that exhibited only graphene. Additionally, we see a set of 3-fold symmetric spots one of which is labeled "hBN", along with a dimmer set of 3-fold spots rotated 60° with respect to the more intense spots. We can confidently ascribe these all six of these spots to the h-BN evident in the LEEM image; based on the diffraction, the h-BN island is seen to be a single crystal (with the expected three-fold, $C_{3v}$ symmetry). Comparing the positions of the h-BN spots to the graphene-derived spots in the pattern, we see they are aligned at 30° relative to the graphene spots.

Examining other h-BN grains in Fig. 2(a), it is apparent that they tend to be oriented in nearly the same manner as the one labelled by "f", or with 180° rotation from that. This preferred orientation, at 30° relative to the EG, is in good agreement with the wide-area LEED pattern of Fig. 1(b). The grains are often found to form at step edges of the underlying SiC, i.e. heterogeneous nucleation, but nevertheless the grains do *not* have a specific orientation relative to the step edges (see Fig. S4 for details). Hence, it appears that the 30° misorientation is an energetic minimum for the system. This conclusion is perhaps surprising, since given the relatively close lattice match (1.8% mismatch) between h-BN and graphene it might be expected that the h-BN would grow with the *same* orientation as the graphene. However, recently Wang et al. have observed two energetically stable orientation for graphene on h-BN, namely, at 0° and 30° misorientation.[41] They argue, based on first-principles computations for a small (1.9 nm) flake of graphene on h-BN, that the former is substantially more stable, by 7.2 meV/atom. With the methodology of Ref. [42], we extend their analysis such that it applies to large flakes, i.e. extending over several unit cells of the moiré pattern (as shown in Fig. S5). In that case we find that the 0° and 30° misorientations have nearly the same energy, within 0.6 meV/atom.

The growth mechanism of the h-BN can be determined by examining AFM results, as presented in Fig. 3(a). Overall, the predominant features are plateaus extending across the image; this morphology is quite common for epitaxial graphene on SiC(0001).[34,35,36,43,44] Islands of h-BN can be identified as the triangles with slightly raised (bright) edges; the corner of one such triangle is indicated by the arrow in Fig. 3(a). The majority of such islands are seen to consist of a thin h-BN layer, consistent with the ML coverage found for most islands in the LEEM images. However, some islands, such as the one located in the lower left-hand corner of Fig. 3(a), appear to consist of many layers in a somewhat complex shape. We interpret those as forming multi-layer h-BN, and indeed, in the LEEM images some islands have a differing contrast with different reflectivity spectra (as shown in Fig. S3), consistent with multi-layer h-BN.

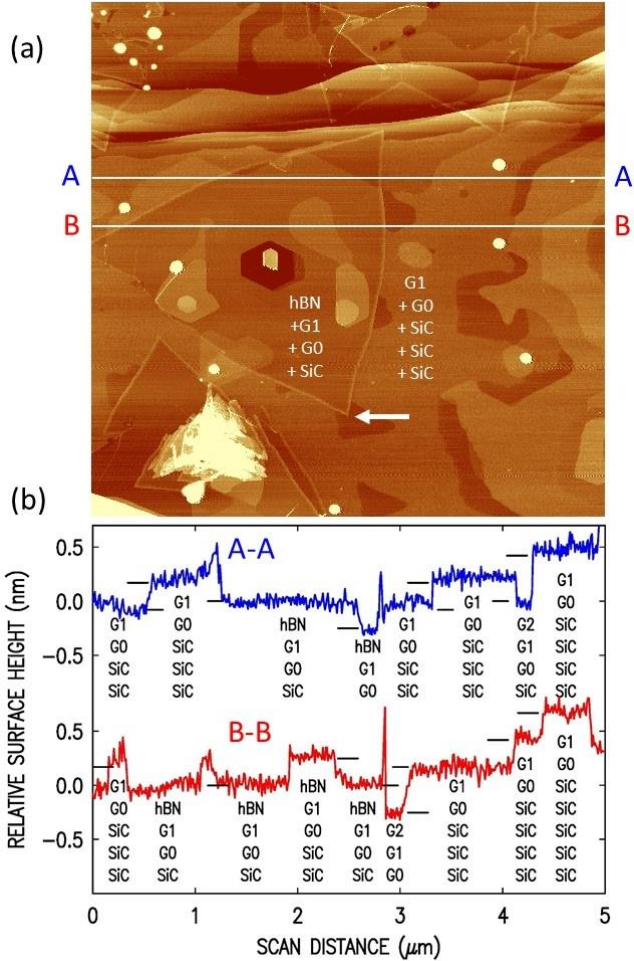

FIG 3. (a) AFM image of sample after h-BN growth at 1100 °C. An arrow indicates the lower corner of a ~2-μm-size triangle with slightly raised (brighter) edges. This triangle and the others seen in the image are identified as h-BN islands. (b) Surface height along the lines A-A and B-B indicated in (a). Thin horizontal lines indicate the expected heights of terraces, according to the labeled arrangement of atomic planes extending from the surface downwards into the SiC.

Close examination of the h-BN islands in Fig. 3(a), e.g. the one whose corner is indicated by the arrow, reveals that, although the edges of the island are slightly raised relative to the surrounding graphene, the body of the triangle actually seems to be *deeper* than the surrounding sample surface. To quantify this effect, line cuts through the image are shown in Fig. 3(b). To understand the height variation across the image, we note that the SiC bilayers of the SiC(0001) substrate are separated by 0.25 nm and separation of the 2D layers (either graphene or h-BN) is 0.33 nm (there is a buffer layer beneath the EG,[34,35,36,45,46,47,48] but it is uniformly present over the entire sample so its separation from neighboring layers does not affect height variations across the surface). Using these interplanar separations, together with the LEER result that the h-BN islands typically contain 1 ML of h-BN on 1 ML of graphene, we label the various heights seen in the line cut according to the number and composition of surface and subsurface planes (i.e. h-BN; graphene layers, $G_1$ and $G_2$; buffer layer, $G_0$; and SiC bilayer(s)). The h-BN island is generally found to contain one additional 2D layer compared to the area immediately surrounding it.

From the plane assignments in Fig. 3(b), it is apparent that there are, in general, two fewer SiC planes below an h-BN island as compared to the surrounding surface area. This reduction in the number of SiC planes accounts for the fact that the h-BN islands in the AFM images appear to be *deeper* than surrounding surface, even though the islands contain an additional 2D layer, i.e. the h-BN layer (see Fig. S6 for statistics on the depth of the h-BN below the surrounding SiC). Hence, we deduce the mechanism of the island formation: Initially the h-BN *substitutes* for the topmost graphene layer (typically G1) on the surface. This substitution liberates C atoms, which are then available to form a new graphene layer *below* the h-BN layer, i.e. at the interface between the existing graphene and the SiC. This formation mechanism for the EG at this interface is well known based on prior studies[34,35,36,44,46,47,48,49] – graphene forms by Si leaving the surface, thereby liberating C atoms to form a new (subsurface) graphene layer. In the absence of additional C atoms produced by the substitution of the h-BN for the top graphene layer, it is necessary to remove Si from 3.1 of the SiC bilayers in order to produce sufficient C to form a new graphene layer.[44] Our data indicates slightly more than 2 SiC bilayers are being consumed by the formation of the new graphene layer. Hence, the remaining amount of C needed for forming the new layer must come from the substitution of the h-BN for the top graphene layer. That is, the amount of C corresponding approximately to what is in 1 SiC bilayer (i.e. 12 atoms/nm$^{-2}$) is supplied by C coming from the substitution. This result is not surprising, since surely not all of the C from the substitution will go directly to the graphene/SiC interface. Rather, these C atoms from the substitution will diffuse to various locations; we find that approximately 1/3 of them are used for forming the new graphene layer, and the remaining ones presumably are incorporated elsewhere (e.g. near SiC step edges, to form additional graphene there) or they leave the surface, e.g. as methane or ethane.

Finally, we comment briefly on the effects of varying the temperature of the h-BN growth on the graphene/SiC surface morphology and the uniformity/crystallinity of the as-grown h-BN. Figure S7 shows results for growth at 950 and 1300 °C. The LEED pattern for the 950 °C growth shows a ring of uniform intensity, at a wavevector consistent with h-BN. AFM reveals a surface covered in an inhomogeneous film, exhibiting a web-like pattern with typical dimension on the 100 nm scale. For this relatively low growth temperature, we conclude that h-BN still forms, but with small domains and no preferred orientation. Perhaps an h-BCN alloy has formed on the surface, as previously observed in several studies.[26,50] For the case of the sample grown at 1300 °C, we find a diffraction pattern with no ring, but with six spots that are rotationally aligned with the SiC spots and an additional six spots that are aligned at 30° to the SiC spots. All of these spots have wavevector consistent with that of h-BN. SiC spots are still apparent in the pattern, but the sixfold satellite spots surrounding the SiC spots are absent, indicating an absence of the $6\sqrt{3}$ buffer layer. LEER spectra of this sample show no features that are characteristic of graphene, and AFM images reveal large changes in the overall surface morphology compared to that of the starting EG surface. We conclude that significant etching of the surface has occurred (i.e. likely due to the H from the borazine), and no graphene nor C-rich buffer layer exists on the surface. We find h-BN with orientations of both 0° and 30° relative to the SiC, with considerably more of the latter compared to the former, judging from the intensity of the LEED spots. LEEM/LEER data indicates that the surface contains two different types of areas, in agreement with AFM, one of which consists of thick, 3-dimensional islands of h-BN and the other of which consists of flat SiC terraces likely terminated by H or N or some other species.

In conclusion, we have demonstrated the growth of h-BN monolayers on epitaxial graphene on SiC(0001), producing relatively large, single-domain h-BN crystals (nearly 2 μm on a side as determined by LEEM and AFM), oriented at 30° relative to the underlying graphene. We find that ML-thick h-BN forms via a substitution reaction with the graphene (as opposed to simple deposition on top of the graphene). We note that this same mechanism has been observed by previous workers,[50] and also in our prior study of h-BN growth on graphene-covered copper substrates.[26] However, in those prior works it was found that h-BCN alloy dominantly formed on the surface, whereas in the present work we obtain pure h-BN formation. The difference between the results arises, we believe, from the ability of the SiC substrate to accommodate the C atoms that are liberated by the substituting B and N, i.e. by forming new graphene planes (although it is necessary in this case for Si atoms to leave the surface). For the metallic substrate of the prior works, the C atoms apparently remain active on the graphene/h-BCN surface, thus inhibiting the formation of large-area, pure h-BN grains.

This work was supported by the Center for Low Energy Systems Technology (LEAST), one of the six SRC STARnet Centers, sponsored by MARCO and DARPA, and by the National Science Foundation, grant DMR-1205275. We are grateful to Devashish Gopalan and Vineetha Bheemarasetty for useful discussions.

# References

[1] A. K. Geim and I. V. Grigorieva, Nature **449**, 419 (2013).

[2] J. Xue, J. Sanchez-Yamagishi, D. Bulmash, P. Jacquod, A. Deshpande, K. Watanabe, T. Taniguchi, P. Jarillo-Herrero, and B. J. LeRoy, NNat. Materials **10**, 282 (2011).

[3] C. R. Dean, A. F. Young, I. Meric, C. Lee, L. Wang, S. Sorgenfrei, K. Watanabe, T. Taniguchi, P. Kim, K. L. Shepard, and J. Hone, Nat. Nanotechnol. **5**, 722 (2010).

[4] L. Britnell, R. V. Gorbachev, A. K. Geim, L. A. Ponomarenko, A. Mishchenko, M. T. Greenaway, T. M. Fromhold, K. S. Novoselov, L. Eaves, Nat. Comm. **4**, 1794 (2013).

[5] T. Roy, M. Tosun, X. Cao, H. Fang, D.-H. Lien, P. Zhao, Y.-Z. Chen, Y.-L. Chueh, J. Guo, A. Javey, ACS Nano **9**, 2071 (2015).

[6] A. Mishchenko, J. S. Tu, Y. Cao, R. V. Gorbachev, J. R. Wallbank, M. T. Greenaway, V. E. Morozov, S. V. Morozov, M. J. Zhu, S. L. Wong, F. Withers, C. R. Woods, Y-J. Kim, K. Watanabe, T. Taniguchi, E. E. Vdovin, O. Makarovsky, T. M. Fromhold, V. I. Fal'ko, A. K. Geim, L. Eaves, and K. S. Novoselov, Nat. Nanotech. **9**, 808 (2014).

[7] B. Fallahazad, K. Lee, S. Kang, J. Xue, S. Larentis, C. Corbet, K. Kim, H. C. P. Movva, T. Taniguchi, K. Watanabe, L. F. Register, S. K. Banerjee, E. Tutuc, Nano Letters **15**, 428 (2015).

[8] X. Li, W. Cai, J. An, S. Kim, J. Nah, D. Yang, R. Piner, A. Velamakanni, I. Jung, E. Tutuc, S. K. Banerjee, L. Colombo, R. Ruoff, Science **324**, 1312 (2009).

[9] L. Gao, J. R. Guest, and N. P. Guisinger, Nano Letters **10**, 3512 (2010).

[10] Q. Yu, J. Lian, S. Siriponglert, H. Li, Y. P. Chen, S.-S. Pei, Appl. Phys. Lett. **93**, 113103 (2008).

[11] Y. Murata, V. Petrova, B. B. Kappes, A. Ebnonnasir I. Petrov, Y.-H. Xie, C. Ciobanu, and S. Kodambaka, ACS Nano **4**, 6509 (2010).

[12] J. Yin, J. Li, Y. Hang, J. Yu, G. Tai, X. Li, Z. Zhang, and W. Guo, Small **12**, 2942 (2016).

[13] S. Caneva, R. S. Weatherup, B. C. Bayer, B. Brennan, S. J. Spencer, K. Mingard, A. Cabrero-Vilatela, C. Baehtz, A. J. Pollard, S. Hofmann, Nano Lett. **15**, 1867 (2015).

[14] Y. Song, C. Zhang, B. Li, G. Ding, D. Jiang, H. Wang, X. Xie, Nanoscale Research Letters **9**, 367 (2014).

[15] J. Yun, H. Oh, J. Jo, H. H. Lee, M. Kim and G.-C. Yi, 2D Mater. **5**, 015021 (2018).

[16] Z. Liu, L. Song, S. Zhao, J. Huang, L. Ma, J. Zhang, J. Lou, P. M. Ajayan, Nano Letters **11**, 2032 (2011).

[17] Y.-J. Cho, A. Summerfield, A. Davies, T. S. Cheng, E. F. Smith, C. J. Mellor, A. N. Khlobystov, C. T. Foxon, L. Eaves, P. H. Beton, and S. V. Novikov, Scientific Reports **6**, 34474 (2016).

[18] M. Heilmann, M. Bashouti, H. Riechert, and J. M. J. Lopes, 2D Mater. **5**, 025004 (2018).

[19] Luxmi, S. Nie, P. J. Fisher, R. M. Feenstra, G. Gu, and Y. Sun, J. Electron. Mat. **38**, 718 (2009).

[20] V. Ramachandran, M. F. Brady, A. R. Smith, R. M. Feenstra, and D. W. Greve, J. Electron. Mat. **27**, 308 (1997).

[21] A. J. Van Bommel, J. E. Crombeen, and A. Van Tooren, Surf. Sci. **48**, 463 (1975).

[22] I. Forbeaux, J.-M. Themlin, and J.-M. Debever, Phys. Rev. B **58**, 16396 (1998).

[23] W. Chen, H. Xu, L. Liu, X. Gao, D. Qi, G. Peng, S. C. Tan, Y. Feng, K. P. Loh, and A. T. S. Wee, Surf. Sci. **596**, 176 (2005).

[24] C. Riedl, A. A. Zakharov, and U. Starke, Appl. Phys. Lett. **93**, 033106 (2008).

[25] P. J. Fisher, Luxmi, N. Srivastava, S. Nie, and R. M. Feenstra, J. Vac. Sci. Technol. A **28**, 958 (2010).

[26] D. P. Gopalan, P. C. Mende, S. C. de la Barrera, S. Dhingra, J. Li, K. Zhang, N. A. Simonson, J. A. Robinson, N. Lu, Q. Wang, M. J. Kim, B. D'Urso, and R. M. Feenstra, J. Mater. Res. **31**, 945 (2016).

[27] P.C. Mende P.C. Mende, Q. Gao, A. Ismach, H. Chou, M. Widom, R. Ruoff, L. Colombo, and R.M. Feenstra, Surf. Sci. **659**, 31 (2017).

[28] J. B. Hannon and R. M. Tromp, *Low-Energy Electron Microscopy for Nanoscale Characterization,* in Handbook of Instrumentation and Techniques for Semiconductor Nanostructure Characterization, ed. R. A. Haight, et al. (World Scientific, Singapore, 2012).

[29] H. Hibino, H. Kageshima, F. Maeda, M. Nagase, Y. Kobayashi, and H. Yamaguchi, Phys. Rev. B **77**, 075413 (2008).

[30] C. M. Orofeo, S. Suzuki, H. Kageshima, and H. Hibino, Nano Research **6**, 335 (2013).

[31] R. M. Feenstra, N. Srivastava, Q. Gao, M. Widom, B. Diaconescu, T. Ohta, G. L. Kellogg, J. T. Robinson, and I. V. Vlassiouk, Phys. Rev. B **87**, 041406(R) (2012).

[32] N. Srivastava, Q. Gao, M. Widom, R. M. Feenstra, S. Nie, K. F. McCarty, and I. V. Vlassiouk, Phys. Rev. B **87**, 245414 (2013).

[33] Q. Gao, P. C. Mende, M. Widom, and R. M. Feenstra, J. Vac. Sci. Technol. **33**, 02B105 (2015).

[34] C. Virojanadara, M. Syväjarvi, R. Yakomova, L. I. Johansson, A. A. Zakharov, and T. Balasubramanian, Phys. Rev. B **78**, 245403 (2008).

[35] K. V. Emtsev, A. Bostwick, K. Horn, J. Jobst, G. L. Kellogg, L. Ley, J. L. McChesney, T. Ohta, S. A. Reshanov, J. Röhrl, E. Rotenberg, A. K. Schmid, D. Waldmann, H. B. Weber, and T. Seyller, Nature Mater. **8**, 203 (2009).

[36] Luxmi, N. Srivastava, R. M. Feenstra, and P. J. Fisher, J. Vac. Sci. Technol. B **28**, C5C1 (2010).

[37] S. Kim, J. Ihm, H. J. Choi, and Y.-W. Son, Phys. Rev. Lett. **100**, 176802 (2008).

[38] K. V. Emstev, F. Speck, Th. Seyller, L. Ley, and J. D. Riley, Phys. Rev. B 77, 155303 (2008).

[39] C. Riedl, C. Coletti, T. Iwasaki, A. A. Zakharov, and U. Starke, Phys. Rev. Lett. **103**, 246804 (2009).

[40] M. Conrad , F. Wang, M. Nevius, K. Jinkins, A. Celis, M. Narayanan Nair, A. Taleb-Ibrahimi, A. Tejeda, Y. Garreau, A. Vlad, A. Coati, P. F. Miceli, and E. H. Conrad, Nano Lett. **17**, 341 (2017).

[41] D. Wang, G. Chen, C. Li, M. Cheng, W. Yang, S. Wu, G. Xie, J. Zhang, J. Zhao, X. Lu, P. Chen, G. Wang, J. Meng, J. Tang, R. Yang, C. He, D. Liu, D. Shi, K. Watanabe, T. Taniguchi, J. Feng, Y. Zhang, and G. Zhang, Phys. Rev. Lett. **116**, 126101 (2016).

[42] J. Jung, A. M. DaSilva, A. H. MacDonald, and S. Adam, Nat. Comm. **6**, 6308 (2015).

[43] N. Srivastava, G. He, Luxmi, P. C. Mende, R. M. Feenstra, and Y. Sun, J. Phys. D: Appl. Phys. **45**, 154001 (2012).

[44] J. B. Hannon and R. M. Tromp, Phys. Rev. B **77**, 241404(R) (2008).

[45] S. Kim, J. Ihm, H. J. Choi, and Y.-W. Son, Phys. Rev. Lett. **100**, 176802 (2008).

[46] K. V. Emstev, F. Speck, Th. Seyller, L. Ley, and J. D. Riley, Phys. Rev. B 77, 155303 (2008).

[47] C. Riedl, C. Coletti, T. Iwasaki, A. A. Zakharov, and U. Starke, Phys. Rev. Lett. **103**, 246804 (2009).

[48] M. Conrad , F. Wang, M. Nevius, K. Jinkins, A. Celis, M. Narayanan Nair, A. Taleb-Ibrahimi, A. Tejeda, Y. Garreau, A. Vlad, A. Coati, P. F. Miceli, and E. H. Conrad, Nano Lett. **17**, 341 (2017).

[49] Luxmi, N. Srivastava, G. He, R. M. Feenstra, and P. J. Fisher, Phys. Rev. B **82**, 235406 (2010).

[50] L. Ci, L. Song, C. Jin, D. Jariwala, D.Wu, Y. Li, A. Srivastava, Z. F.Wang, K. Storr, L. Balicas, Feng Liu and P. M. Ajayan, Nature Mater. **9**, 430 (2010).

**Substitutional mechanism for growth of hexagonal boron nitride on epitaxial graphene - SUPPLEMENTARY MATERIAL**

Patrick C. Mende, Jun Li, and Randall M. Feenstra
Dept. Physics, Carnegie Mellon University, Pittsburgh, PA 15213

**Growth Procedure:** Growth was performed in a home-built vacuum system with base pressure of $1\times10^{-9}$ Torr, employing a resistively heated graphite heater strip and a disappearing filament pyrometer for measuring temperature (with accuracy of ±50 °C).[1] SiC(0001) samples (1 cm × 1 cm, Si-face, epi-ready) were first hydrogen-etched to improve their surface morphology,[2] and then graphene was grown by high-temperature sublimation of silicon in 1 atm of Ar.[3,4,5] The samples were characterized by LEED in a connected vacuum chamber, and then returned to the growth chamber where borazine, $(BH)_3(NH)_3$, was introduced until the desired pressure was obtained. The sample temperature was then rapidly (within ~10 s) increased, and after a growth time of 30 minutes the sample temperature was ramped to room temperature over the course of 5 minutes. Following all h-BN growth runs, the graphite heater strip was hydrogen etched in 1 atm of hydrogen at 1700°C for 5 min. This process was found to be necessary to remove BN from the heater strip, and hence prevent unintentional BN deposition (from the heater strip itself) on a subsequent sample.

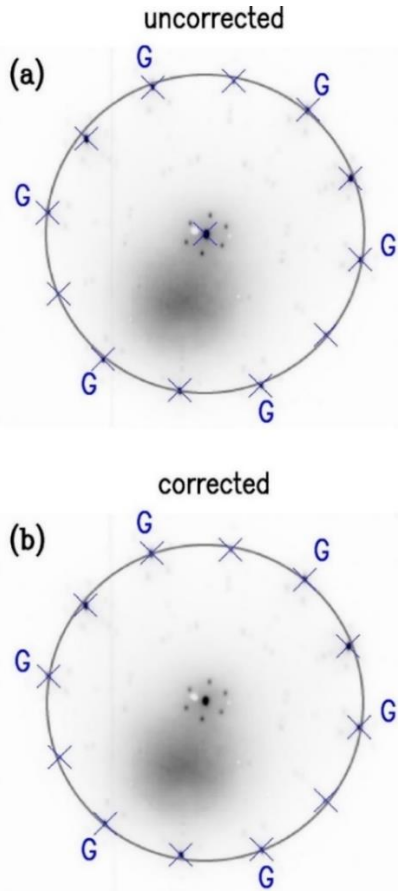

**Additional Data Figures:**

FIG S1. (a) μLEED pattern from Fig. 2(f) of main text, with circle indicating approximate location of graphene (G) spots. Some of these spots are seen to be outside the circle, and others inside, indicating a slight distortion of the pattern (i.e. due to the electron optics of the LEEM). (b) Corrected μLEED pattern in which the pattern from (a) is compressed by 2% along a line with angular orientation of 39° CCW from the horizontal. The resultant graphene spots are seen to have better positioning relative to the circle, and similarly for the h-BN spots. Measuring separations between pairs of spots that span the circle, and computing averages with standard deviations, we obtain a difference between the wavevector magnitudes of the graphene and h-BN spots of 1.97 ± 0.79% for the uncorrected pattern, and 1.89 ± 0.13% for the corrected pattern. The correction is thus seen to significantly reduce the error range in the result, but without significantly affecting the midpoint (mean) value.

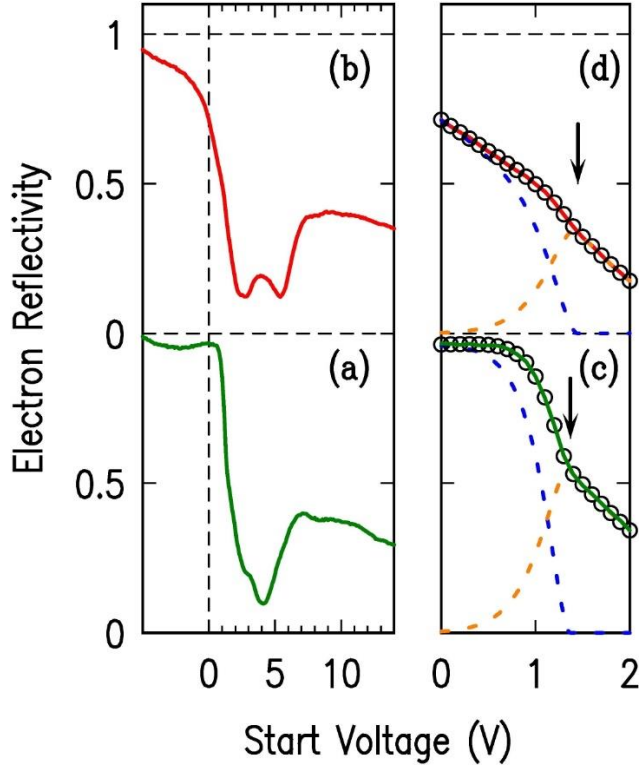

FIG S2. (a) and (b) LEER spectra, from Fig. 2(c) of main text, spectra A and B, respectively, plotted as a function of the start voltage (which is the voltage on sample relative to that on the LaB$_6$ thermionic cathode in the LEEM). (c) and (d) Details of fitting to the spectra, to determine the onset voltages (arrows). At the onset voltage, $V_0$, the vacuum levels of the sample and the cathode are aligned, i.e. the work function difference between sample and LaB$_6$ cathode is given by $\Delta W = eV_0$. From the fitting procedure (described in detail in Ref. [6]), the onset voltages for the two cases shown are determined to be 1.37 ± 0.01 and 1.45 ± 0.05 V, respectively. For spectra C and D of Fig. 2(c), the onsets are found to be 1.43 ± 0.01 and 1.40 ± 0.04 V, respectively. For each case, the maximum in the distribution of electrons emitted by the cathode occurs at an energy $\sigma_c$ above its vacuum level, where $\sigma_c$ is width of the thermionic distribution which is also obtained from the fitting procedure (0.18 eV for the present data). Hence, to place the observed spectra on a scale corresponding to the energy of a sample state relative to the vacuum level (at that location on the sample surface), the spectra are shifted by $-\Delta W + \sigma_c$.

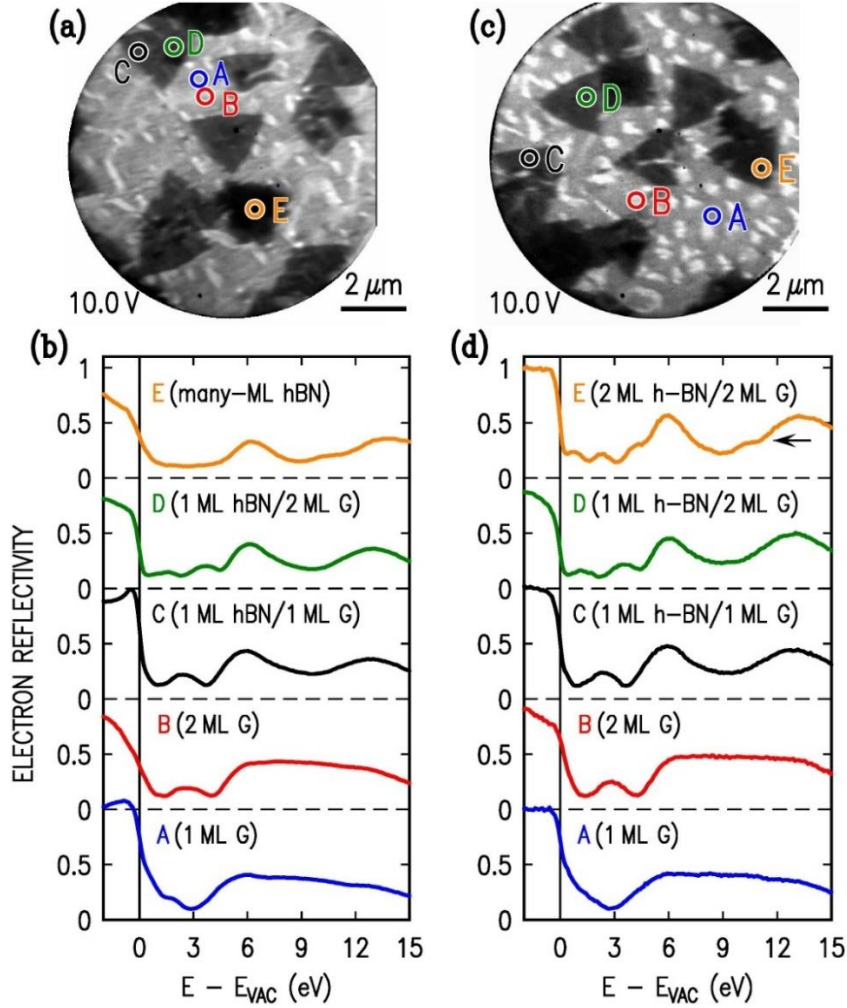

FIG S3. (a) and (c) LEEM images from two different areas of sample grown at 1100 °C, acquired at start voltage of 10 V (image (a) is same as Fig. 2(a) of main text). (b) and (d) LEER spectra acquired from the locations indicated in (a) and (c), respectively. Spectra are identified as originating from the numbers of h-BN and graphene (G) layers as listed. In spectrum E of panel (d), an arrow indicates a shoulder in the reflectivity, indicative of h-BN with 2 ML thickness. Spectrum E of panel (b) shows a flat-bottomed structure for energies of 1 – 4 eV, indicative of many MLs of h-BN.

Interpretation of the LEER spectra is made as described in the main text: In the low-energy range of 0 – 6 eV, oscillations in the LEER spectra arise from *interlayer states* of the h-BN and graphene,[7,8,9] so the number of spectral minima equals the number of h-BN and/or graphene layers. For the broad minimum seen in some of the LEER spectra at 8 – 12 eV, this feature has been identified as being derived from states of the h-BN that are localized *on* the atomic planes.[10] Such *intralayer states* are forbidden, by symmetry, to produce reflectivity minima for graphene,[10] so that this minimum at ~10 eV allows us to distinguish between h-BN and graphene. This interpretation of the LEER spectra has been firmly established in prior studies of h-BN and graphene on Cu and Ni substrates.[6,8,10] For example, 2 ML of h-BN on Ni produces spectrum A in Fig. 1 of Ref. [10] showing a single minimum in the 0 – 6 eV range as well as a minimum at ~10 eV; the latter arises from the intralayer states of the h-BN whereas the former is the interlayer state between the 2 h-BN layers. Only a single interlayer state forms in this case because the space between h-BN and Ni is much too small to permit an interlayer to form there, as revealed by explicit, first-principles simulation of the spectra (although the same simulations demonstrate that an additional, small feature observed near 0 eV in those spectra arises from a Shockley state associated with the Ni surface).[10]

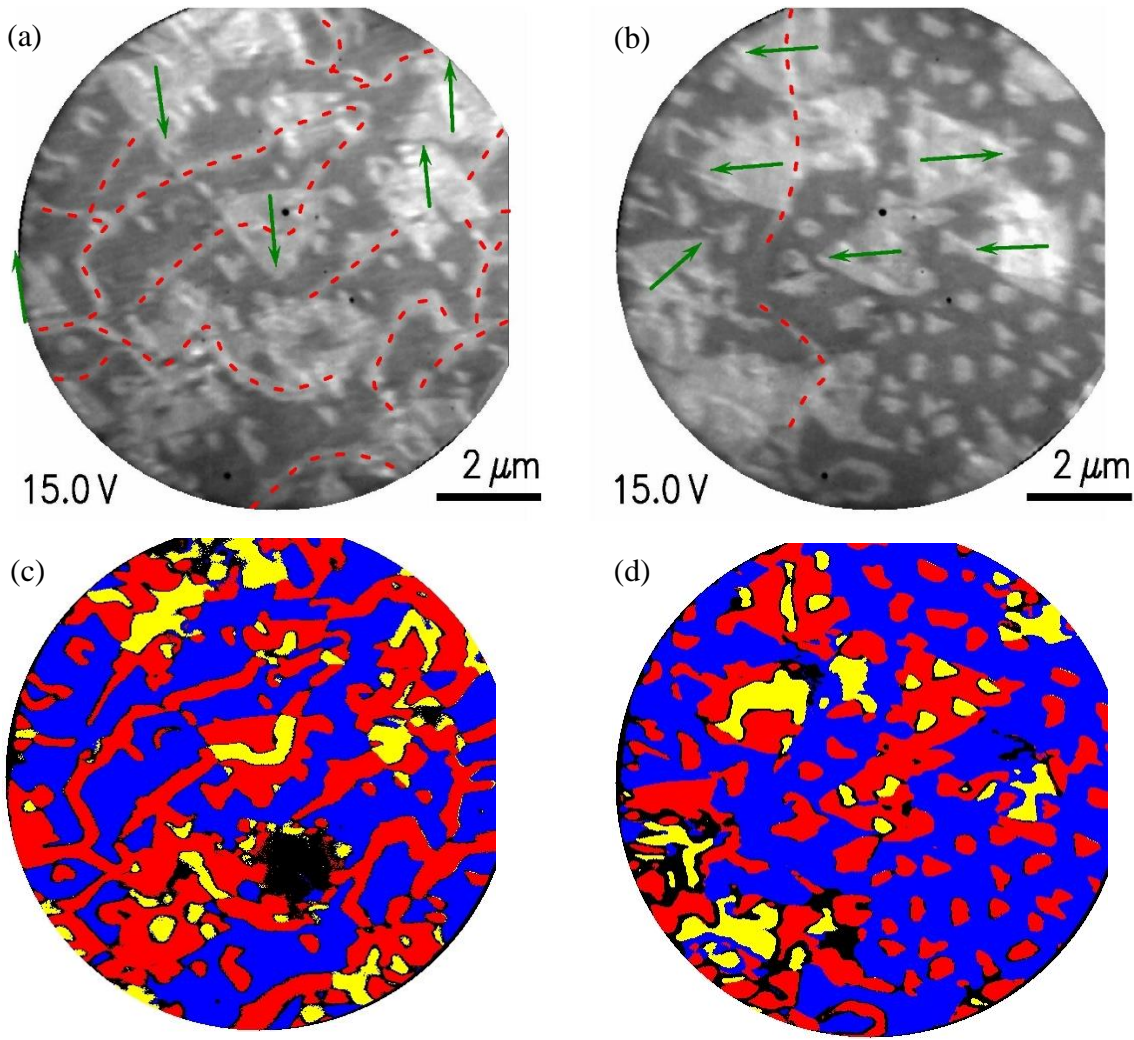

FIG S4. (a,b) LEEM images acquired at 15 V, and (c,d) thickness mapping (blue, red, yellow correspond to 1, 2, or 3 ML of graphene or h-BN, and black corresponds to an ill-defined reflectivity curve, e.g. near a domain boundary or for a flat-bottomed reflectivity minimum such as spectrum E of Fig. S3(b)). Images (a,c) and (b,d) are obtained from the same surface areas as Figs. S3(a) and (c), respectively; these are two different areas of the same sample, but imaged during separate LEEM sessions (there was a 90° rotation in the mounting of the sample between the two sessions). Step edges, as revealed by the bright strips of 2-ML-thick graphene in (a,b), or by the red-on-blue or yellow-on-red strips in (c,d), are indicated by red dashed lines in (a,b), whereas the orientations of h-BN islands are indicated by green arrows. In (a), the h-BN islands are seen to be generally pointing along the vertical, whereas in (b) they are pointing along the horizontal. In both cases, the islands do not appear to specifically follow the step directions, which themselves are rather serpentine. Note that several islands on the right-hand side of images (b,d) do *not* appear to be nucleated at step edges, but nevertheless maintain the same orientation as the islands in the upper left-hand portion of the image that *are* nucleated at a step edge.

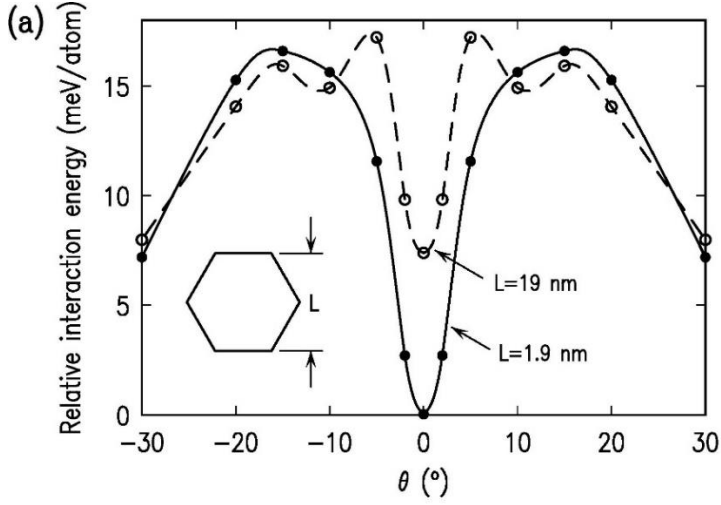

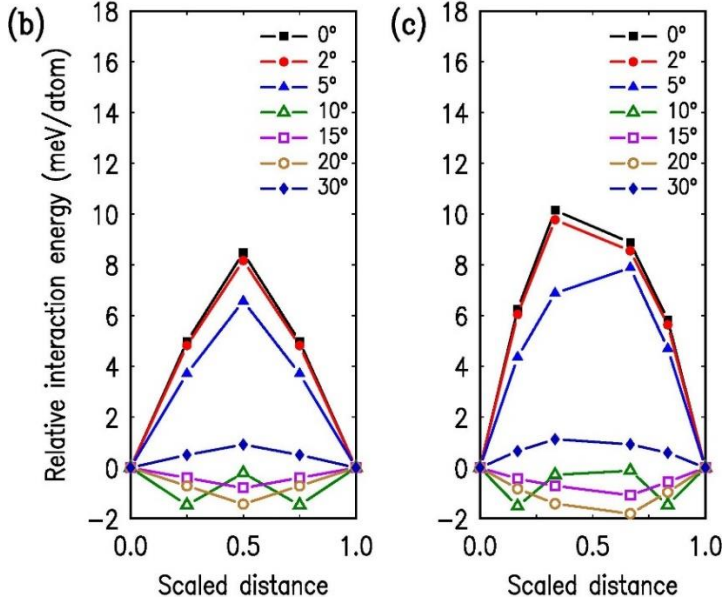

FIG S5. (a) Relative interaction energy between a hexagonal flake of graphene on h-BN, for flake sizes of 1.9 (solid line) and 19 (dashed line) nm, as a function of misorientation angle between the graphene and h-BN. Results for the 1.9-nm flake are fit to the first-principles results of Ref. [11], with the extension to the larger flake made using the method of Ref [12]. For the small flake at small misorientation angles, many of its C atoms are located at energetically favorable locations above the h-BN (i.e. with half of the C atoms nearly above N atoms), and hence the energy is relatively low. For the larger flake, a much greater variety of sites for the C atoms occur, and hence the interaction energy increases.

(b,c) Relative interaction energy as a function of translational shift of the 1.9-nm hexagonal flake relative to the h-BN, with the shift occurring in the zigzag and armchair directions, respectively. Again, results are fit to the first-principles results of Ref. [11], utilizing the expression for the interaction energy of graphene on h-BN as formulated in Ref. [12]. There, the interaction energy of a single unit cell of graphene on h-BN as a function of its position $\mathbf{r} \equiv (x, y)$ relative to an h-BN unit cell is a periodic function (i.e. period of h-BN) expressed as a sum over reciprocal lattice vectors, $V_{12}(\mathbf{r}) = \sum_{\mathbf{G}} c_{\mathbf{G}} e^{i\mathbf{G}\cdot\mathbf{r}}$, where we employ the first three rings of $\mathbf{G}$ vectors (the second and third rings make only a small contribution to our results). In each ring there are three $\mathbf{G}$ vectors with the same $c_{\mathbf{G}}$ values, and three with the complex conjugate of that. The energy of an entire flake of graphene over h-BN is then given by $\frac{1}{N}\sum_{i,j} V_{12}(\mathbf{r}_{i,j})$ where $\mathbf{r}_{i,j}$ are the locations of the graphene unit cells in the flake and $N$ is the number of graphene unit cells in the flake. We evaluate this sum for the 1.9-nm flake, and choose the values of the coefficients such that panels (b,c) and the solid line of (a) are a good fit to the results of Ref. [11] (see their Figs. 3(h), S3(b,c); the 25° misorientation angle of their Fig. 3(h) is not considered here, since it's not included in their Figs. S3(b,c)). We then consider the larger, 19-nm flake, thus obtaining the dashed line of (a).

As an explicit illustration of our analysis procedure, let us consider a misorientation of $\theta = 0°$, treated using only the first ring of **G** vectors. Primitive lattice vectors of h-BN are given by $\mathbf{a}_1 = a\mathbf{e}_x$ and $\mathbf{a}_2 = a\left(-\frac{1}{2}\mathbf{e}_x + \frac{\sqrt{3}}{2}\mathbf{e}_y\right)$ where $\mathbf{e}_x$ and $\mathbf{e}_y$ are unit vectors in Cartesian directions, and $a$ is the h-BN lattice constant. Reciprocal lattice vectors are given by $\mathbf{G}_{10} = -\mathbf{G}_{\bar{1}0} = \frac{4\pi}{\sqrt{3}a}\left(\frac{\sqrt{3}}{2}\mathbf{e}_x + \frac{1}{2}\mathbf{e}_y\right)$, $\mathbf{G}_{0\bar{1}} = -\mathbf{G}_{01} = -\frac{4\pi}{\sqrt{3}a}\mathbf{e}_y$, and $\mathbf{G}_{\bar{1}1} = -\mathbf{G}_{1\bar{1}} = \frac{4\pi}{\sqrt{3}a}\left(-\frac{\sqrt{3}}{2}\mathbf{e}_x + \frac{1}{2}\mathbf{e}_y\right)$. The interaction energy between a single unit cell of graphene on h-BN is written as $V_{12}(\mathbf{r}) = \sum_\mathbf{G} c_\mathbf{G} e^{i\mathbf{G}\cdot\mathbf{r}}$. This sum includes a $\mathbf{G} = (0,0)$ term with real coefficient $c_{00}$, as well as the terms from the **G** vectors, with complex coefficients $c_{10} = c_{0\bar{1}} = c_{\bar{1}1} = (c_{\bar{1}0})^* = (c_{01})^* = (c_{1\bar{1}})^*$ such that $V_{12}(\mathbf{r})$ is a real, periodic function. We then consider a flake of graphene on h-BN, containing $N$ graphene unit cells. The interaction energy of the entire flake, per graphene unit cell, is given by $\frac{1}{N}\sum_{i,j} V_{12}(\mathbf{r}_{i,j})$ where $\mathbf{r}_{i,j}$ are the locations of the graphene unit cells in the flake with respect to the h-BN lattice, and the indices $i,j$ label the unit cells. Since $V_{12}(\mathbf{r})$ is periodic in the h-BN lattice period, we only need consider the *difference* in the position of the graphene unit cells relative to the underlying h-BN, $\frac{1}{N}\sum_{i,j} V_{12}(\Delta\mathbf{r}_{i,j})$. Both the h-BN and the graphene unit cells form hexagonal arrays; the positions of the h-BN unit cells can be expressed as $a\left(i - \frac{1}{2}j\right)\mathbf{e}_x + a\frac{\sqrt{3}}{2}j\mathbf{e}_y$ in terms of indices $i,j$. The graphene lattice unit cells are smaller than those of h-BN by a fractional amount $\delta = -0.018$, so that we have $\Delta\mathbf{r}_{i,j} = \left[x_0 + a\delta\left(i - \frac{1}{2}j\right)\right]\mathbf{e}_x + \left[y_0 + a\delta\frac{\sqrt{3}}{2}j\right]\mathbf{e}_y$, where $(x_0, y_0)$ defines the translational alignment between the graphene and h-BN lattices at the center of the flake, $i = j = 0$. We define a hexagonal shape for the flake by placing limits on the $i,j$ values: for $-n \leq j \leq n$ we have $-n \leq i \leq n$ for $j = 0$, $-n + j \leq i \leq n$ for $j > 0$, $-n \leq i \leq n + j$ for $j < 0$, and the total number of unit cells in the flake is $N = 3n^2 + 3n + 1$.

With these definitions, the total energy $\frac{1}{N}\sum_{i,j} V_{12}(\Delta\mathbf{r}_{i,j})$ can be easily evaluated, depending only on the values of $n$, $c_{00}$, $c_{10}$. The latter is a complex constant, which we write in terms of its magnitude $|c_{10}|$ and its complex phase $\arg(c_{10})$. We evaluate these parameters by considering the *variation* in $\frac{1}{N}\sum_{i,j} V_{12}(\Delta\mathbf{r}_{i,j})$, i.e. with no dependence on $c_{00}$, for a 1.9-nm flake ($n = 4$) as $(x_0, y_0)$ varies along the zigzag and armchair directions of the h-BN lattice. This result is shown in Figs. S5(b,c). By choosing values of $|c_{10}| = 1.04$ meV and $\arg(c_{10}) = 187°$, we obtain a good fit between these variations and those from the first-principles computations of Ref. [11] (as mentioned above, we also employ the second and third rings of G vectors in order to slightly improve the match with the first-principles results). Then, to evaluate $c_{00}$, we consider the magnitude of the interaction energy for the 1.9-nm flake for $x_0 = y_0 = 0$, as given in Ref. [11] and shown by the solid line in Fig. S5(a), leading to $c_{00} = 6.75$ meV. Finally, to achieve our goal of evaluating the interaction energy of a larger, 19-nm flake of graphene on h-BN, we simply change the $n$ value defining the boundary of the flake, to $n = 40$. We thus obtain the result shown by the dashed line in Fig. S5(a), and our analysis procedure is repeated at each of the misorientation angles shown there.

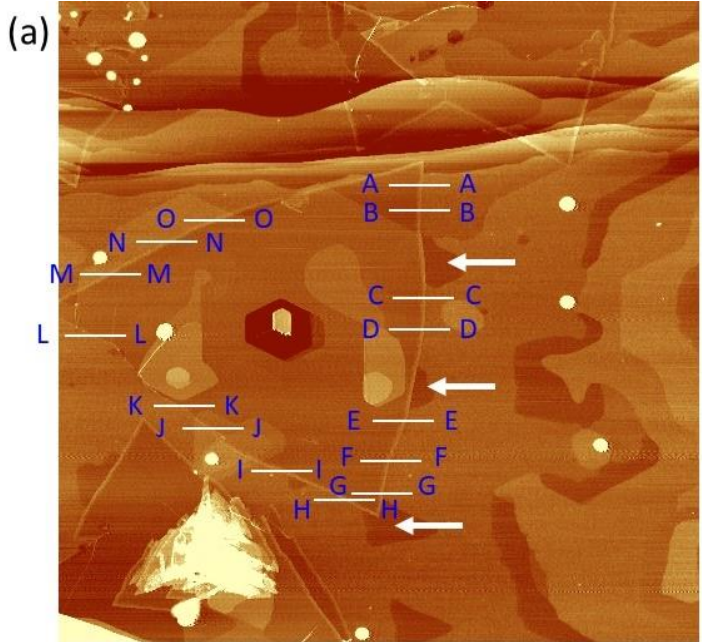
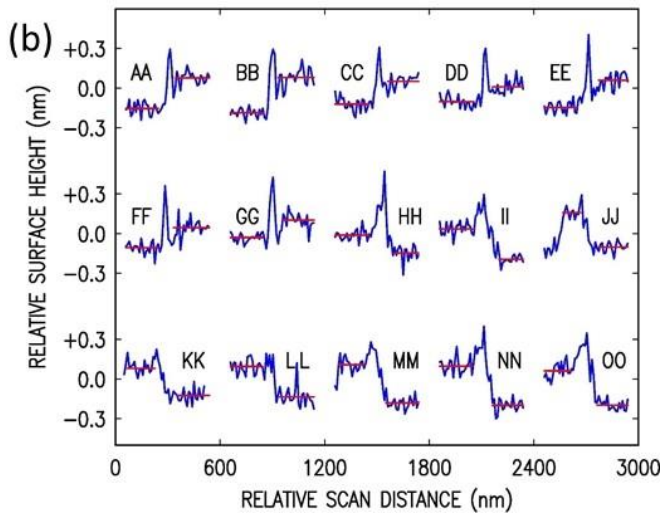
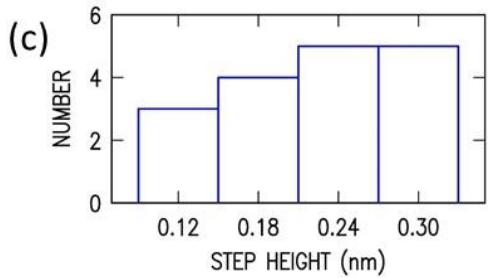

FIG S6. (a) AFM image (5×5 μm$^2$) showing line cuts along the border of an h-BN island. (b) Height variation along the line cuts, with red lines showing portions of the line cuts used to compute terrace heights. (c) Histogram of results for step height between h-BN island and surrounding SiC areas. In choosing the locations of the line cuts, areas marked by the white arrows in (a) are avoided, at which the h-BN island is seen to be *higher* than the SiC. To understand such areas, we refer to the thickness map of Fig. 2(d), where similar-shaped small areas are seen extending out from the h-BN triangular island. These areas are found to have the *same* number of 2D layers as the island (that is, small red areas extending out from the edge of the red h-BN island in Fig. 2(d)), and hence we interpret the areas in (a) as similarly having the *same* number of 2D layers as the h-BN island. (Also see cut BB of Fig. 3(b), where one such area is explicitly labelled, just to the right of the h-BN island). All other areas of the SiC surrounding the h-BN island are found in the thickness map to have one *fewer* 2D plane in the surrounding SiC than in the h-BN island itself. The histogram in (c) yields a mean step height of 0.211 nm and standard deviation of $\sigma = 0.060$ nm. Within $\pm\sigma$, this measured step height is consistent with having two additional SiC bilayers in the SiC area surrounding the h-BN island compared to beneath the island itself, corresponding to a step height of (2×0.25) – 0.33 = 0.17 nm. More quantitatively, we can utilize the error on the mean for the $N = 15$ measurement together with a 95% confidence interval, yielding an uncertainty of $\pm 2\sigma/\sqrt{N} = \pm 0.031$ nm. The lower end of this range, 0.211 – 0.031 = 0.180 nm, is slightly larger, by 0.01 nm, than the expected value of 0.17 nm. However, this small discrepancy is well within the bounds of systematic error in AFM imaging of dissimilar materials.[13]

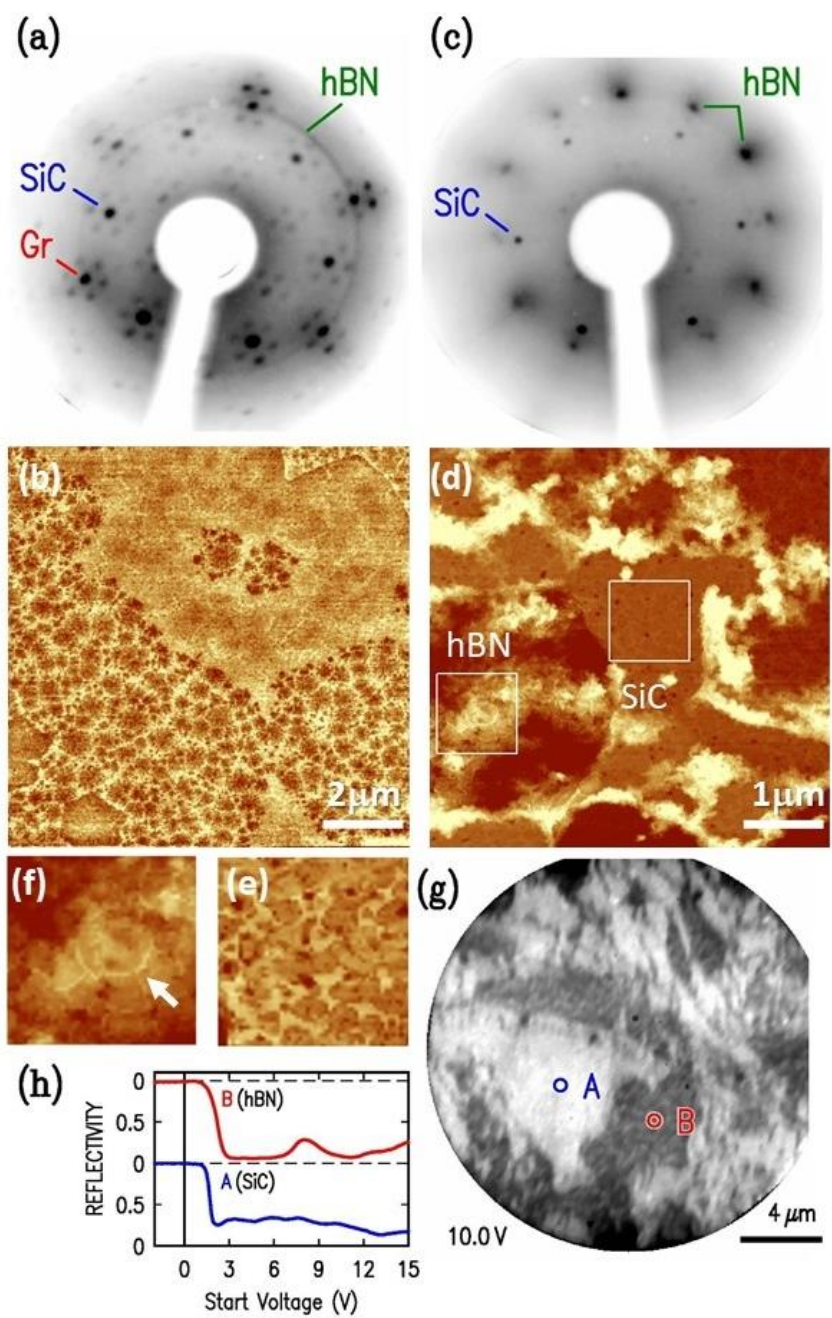

FIG S7. Results for h-BN growth on EG at temperatures of (a,b) 950 °C, and (c-h) 1300 °C.

(a) and (c) LEED patterns, acquired at electron energy of 100 eV;

(b) and (d) AFM images;

(e) and (f) expanded views of the areas shown by white squares in (d). The former appears as a relatively flat surface area, with distinctive "scaly" like appearance. The latter shows a relatively rough, 3-dimensional type of growth, but with occasional distinctive lines that appear to follow crystallographic directions (as indicated by the white arrow);

(g) LEEM image;

(h) LEER spectra extracted from the locations indicated in (g).

For the 1300 °C growth, the surface morphology is greatly changed compared to the starting EG, indicative of etching of the surface. LEER spectrum B indicates thick h-BN (compare with spectrum E of Fig. S3(b)), whereas LEER spectrum A is clearly indicative *not* of h-BN nor of graphene, and we believe that it arises from some sort of SiC surface that may be terminated e.g. by H or N or some other species. Hence, it appears that the graphene on the surface has been etched away, with much of the surface consisting of flat SiC terraces and other parts of the surface being covered with thick, 3-dimensional h-BN islands.

# References

[1] Luxmi, S. Nie, P. J. Fisher, R. M. Feenstra, G. Gu, and Y. Sun, J. Electron. Mat. **38**, 718 (2009).

[2] V. Ramachandran, M. F. Brady, A. R. Smith, R. M. Feenstra, and D. W. Greve, J. Electron. Mater. **27**, 308 (1997).

[3] C. Virojanadara, M. Syväjarvi, R. Yakomova, L. I. Johansson, A. A. Zakharov, and T. Balasubramanian, Phys. Rev. B **78**, 245403 (2008).

[4] K. V. Emtsev, A. Bostwick, K. Horn, J. Jobst, G. L. Kellogg, L. Ley, J. L. McChesney, T. Ohta, S. A. Reshanov, J. Röhrl, E. Rotenberg, A. K. Schmid, D. Waldmann, H. B. Weber, and T. Seyller, Nature Mater. **8**, 203 (2009).

[5] Luxmi, N. Srivastava, R. M. Feenstra, and P. J. Fisher, J. Vac. Sci. Technol. B **28**, C5C1 (2010).

[6] D. P. Gopalan, P. C. Mende, S. C. de la Barrera, S. Dhingra, J. Li, K. Zhang, N. A. Simonson, J. A. Robinson, N. Lu, Q. Wang, M. J. Kim, B. D'Urso, and R. M. Feenstra, J. Mater. Res. **31**, 945 (2016).

[7] R. M. Feenstra, N. Srivastava, Q. Gao, M. Widom, B. Diaconescu, T. Ohta, G. L. Kellogg, J. T. Robinson, and I. V. Vlassiouk, Phys. Rev. B **87**, 041406(R) (2012).

[8] N. Srivastava, Q. Gao, M. Widom, R. M. Feenstra, S. Nie, K. F. McCarty, and I. V. Vlassiouk, Phys. Rev. B **87**, 245414 (2013).

[9] Q. Gao, P. C. Mende, M. Widom, and R. M. Feenstra, J. Vac. Sci. Technol. **33**, 02B105 (2015).

[10] P.C. Mende P.C. Mende, Q. Gao, A. Ismach, H. Chou, M. Widom, R. Ruoff, L. Colombo, and R.M. Feenstra, Surf. Sci. **659**, 31 (2017).

[11] D. Wang, G. Chen, C. Li, M. Cheng, W. Yang, S. Wu, G. Xie, J. Zhang, J. Zhao, X. Lu, P. Chen, G. Wang, J. Meng, J. Tang, R. Yang, C. He, D. Liu, D. Shi, K. Watanabe, T. Taniguchi, J. Feng, Y. Zhang, and G. Zhang, Phys. Rev. Lett. **116**, 126101 (2016).

[12] J. Jung, A. M. DaSilva, A. H. MacDonald, and S. Adam, Nat. Comm. **6**, 6308 (2015).

[13] K. Godin, C. Cupo, and E.-H. Yang, Sci. Reports **7**, 7798 (2017).